# Steady-state distributions of carrier concentration and recombination rates in a solar cell under operating conditions

Isshin Sumiyoshi,[1,*] Yoshitaro Nose,[1,†]

[1]Department of Materials Science and Engineering, Kyoto University, Sakyo-ku, Kyoto 606-8501, Japan

**ABSTRACT.** The steady-state distribution of carrier concentrations in a solar cell under operating conditions is a key source of carrier recombination and directly influences the output current density. In this study, we investigated the effects of illumination and bias voltage on the steady-state distributions of carrier concentrations and recombination rates in a homo-pn junction solar cell using one-dimensional numerical simulations to explore passivation strategies driven by the reduction of carrier concentrations. Simulations under open circuit conditions revealed that controlling the standard states of carriers can enhance the open circuit voltage without changing the carrier concentration. Furthermore, distributing recombination rates conventionally concentrated in the vicinity of the interface into the bulk region, such as within the absorbing layer, improves the open circuit voltage. Our results also showed that changing the carrier distribution in the dark to that under illumination, in other words, transition from an equilibrium to a steady state, is governed by drift rather than diffusion. This means that the electric field, which induces drift, is a primary driving force for carrier separation. Consequently, by optimizing the electric field distribution depending on properties of recombination-causing defects, a higher short circuit current density can be achieved, even without chemical passivation of defects. This work offers fundamental insights into reduction of steady-state recombination rates through precise control of carrier concentration, chemical potential, and electric field distributions.

## I. INTRODUCTION

In optoelectronic devices such as solar cells, suppression of the recombination of photogenerated electron-hole pairs is crucial for enhancing the performance. Carrier recombination directly leads to performance losses. Addressing this issue requires innovative approaches in device design and fabrication. In crystalline silicon (c-Si) solar cells, which are the most efficient and most widespread solar cells, several strategies have been developed to mitigate recombination losses. These strategies include the use of an aluminum back surface field (Al-BSF) [1–5], passivated emitter rear contact (PERC) [6–9], integrated back contact (IBC) [10–14], heterojunction solar cells (HJT) [9,13,15–17], and tunnel oxide passivated contact (TOPCon) [18–20]. Since the technology for growing high–quality silicon single crystals has matured and the impact of bulk recombination has been reduced due in part to the indirect bandgap, optimizing surface passivation has become a key focus for enhancing cell efficiency.

There are two primary strategies for reducing recombination at interfaces. The first strategy is to decrease the density of defects that mediate recombination and act as recombination centers. Dangling bonds at the surface of silicon, which are the prominent defects in c-Si cells, have been primarily passivated by hydrogen [21]. This approach is applied to a hydrogenated amorphous silicon nitride (SiNx) layer on p-type silicon [22,23] and a hydrogenated amorphous silicon (a-Si:H) layer for HJT [24,25]. Modifying the chemical bonds associated with defects prevents them from being ionized or neutralized during solar cell operation, leading to the reduction of recombination. This solution is thus referred to as "chemical passivation". The second strategy is to reduce the minority carrier concentration at the interfaces, which is a source of recombination on the basis of the Schockley-Read Hall (SRH) statistics [26,27]. Since 1963, when M. Wolf predicted the reduction of carrier concentration at the surface of silicon by forming an electric field at the interface [1], theoretical and experimental studies have been reported. The electric field is formed by the combination of Al printing on a silicon back surface and the subsequent alloying process in the Al-BSF technique [1–5] and by depositing a dielectric thin layer in TOPCon cells [18–20], which have successfully reduced recombination and are the core technologies at present. This method is known as "field-effect passivation". Building upon these successful passivation strategies in c-Si solar cells, extending these methods to other semiconductor materials presents a promising avenue for improving solar cell performance across a broader range of technologies.

When applying passivation technologies to various material systems, it is crucial to focus on the intrinsic differences with c-Si cells. In particular, bulk recombination cannot be neglected, especially in direct bandgap semiconductors. It is reasonable that significant efforts are being made to control point defects and to explore recombination centers in absorbing layers, both of which are essential aspects of chemical passivation. First-principles calculations have recently provided meaningful assistance for point defect investigations, while enhancing carrier lifetime by controlling defects remains challenging. For field-effect passivation, in contrast, more feasible methods such as doping and depositing layers are available, and indeed there have been many reports on the back surface field (BSF) [28–32]. However, the BSF suppresses interface recombination at the back electrode, and it is not effective in reducing recombination within semiconductor layers. In this context, extending the consideration of minority carrier concentration beyond just the surface, as is the case with silicon, to throughout the solar cell makes it feasible to achieve the same effect as field-effect passivation. As described above, field-effect passivation contributes to the reduction of minority carrier concentration. Therefore, it is essential to consider the steady-state distribution of carrier concentrations in a solar cell under operating conditions.

*Contact author: sumiyoshi.isshin.35a@st.kyoto-u.ac.jp
†Contact author: nose.yoshitaro.5e@kyoto-u.ac.jp

Despite efforts in the study of c-Si solar cells, several fundamental challenges remain. One major issue is that illumination and applied bias voltage affect carrier distribution differently. To improve power conversion efficiency, both effects must be considered. For c-Si solar cells, discussions about minority carrier concentration, even at the interface, have mostly focused on illumination, with less attention given to bias voltage effects. Another challenge is that the pseudo-Fermi level makes it difficult to understand how drift and diffusion influence carrier distribution. The quasi-Fermi level incorporates both the chemical and electric potential, which reflect carrier concentration and charge potential, respectively. Without clearly separating these, the effects of drift and diffusion on carrier distributions can be confusing. This may explain why discussions about carrier separation are still open for debate.

In this study, we investigated the influence of illumination and bias voltage on carrier distributions in a homo-pn junction solar cell, with the aim of achieving suppression of recombination by reduction of carrier concentrations, using a one-dimensional device simulator, SCAPS-1D [33]. We found that improvement of open circuit voltage can be achieved by distributing the minority carrier concentration under an open circuit into the bulk layer while maintaining the total recombination equal to the total generation. Additionally, the minority carrier distribution under a short circuit is determined by the electric field and optimizing it according to the defect distribution is necessary. Focusing on the steady-state distributions of carrier concentration and recombination provides deep insight into enhancement of solar cell performance without altering the defect characteristics.

## II. GENERAL PRINCIPLE

In this section, we present expressions for discussion on suppressing recombination and improving photovoltaic performance from the perspective of carrier distributions and recombination rates. We work in a hypothetical one-dimensional solar cell with a homo-$pn$ junction assuming that the degradation of solar cell performance due to series and shunt resistance is sufficiently small. The conservation of carriers in a solar cell irradiated by light leads to the following equation,

$$g = j + j_d = r \qquad (1)$$

where $j$ is the net current density, $g$ and $r$ are the integrals of the generation rate $G(x)$, and recombination rate $U(x)$, given by

$$g = q\int G(x)dx \qquad (2)$$
$$r = q\int U(x)dx \qquad (3)$$

respectively. $j_d$ represents the diffusion current density which is electron (or hole) current flowing towards the $p$-type (or $n$-type) layer. The performance of a solar cell is usually evaluated via the response of $j$ to the illumination and the applied bias voltage. Generally, $G(x)$ depends on the absorption coefficient of each layer in a solar cell, and $g$ can be treated as a constant. $j_d$ is negligible in Eq. (1) for the considered solar cells that consist of an absorbing layer with a direct bandgap. In this paper, we focus on $r$ rather than $j$ itself in order to evaluate $j$ using Eq. (1). For that purpose, we investigate distributions of electron concentration in the conduction band, $n_e(x)$, and hole concentration in the valence band, $n_h(x)$, since $r$ is a function of these. Although some reports argue that carrier conductivity, which is the product of mobility and carrier concentration, is the most essential parameter for carrier separation [34,35], it is not a fundamental variable for recombination rates. It is therefore essential to investigate how the carrier distribution behaves in response to light illumination and applied bias voltage.

The steady-state carrier distribution in solar cells under illumination and zero-biased differs from that in the dark due to photogeneration of carriers. The short circuit current density $J_{SC}$, is given by the difference in $g$ and $r$ under this condition as shown in Eq. (1). $g$ is a constant here, and $r$ is determined by the steady-state carrier concentrations. The transient from carrier distribution in the dark to that under illumination is basically considered by the Boltzmann transport equation. However, as explained in Sec. I, an understanding of this transient, especially in the bulk region, remains an open question. In the case of a basic homo-*pn* junction solar cell, the quantum carrier transport is negligible, and the transient is dominated by drift or diffusion. To elucidate their contributions, it is necessary to distinguish between the electric potential gradient (or electric field) and the chemical potential (or carrier concentration) gradient, which are the driving forces of drift and diffusion, respectively.

Conversely, it is commonly understood that when a bias voltage is applied to a solar cell, the chemical potentials of electrons and holes are altered. These chemical potentials are defined by

$$\mu_e = E_C + kT \ln \frac{n_e}{N_C} \qquad (4)$$

$$\mu_e = E_V + kT \ln \frac{n_h}{N_V} \qquad (5)$$

where $k$ is the Boltzmann constant, and $T$ is the absolute temperature. $E_i$ and $N_i$ ($i$ = C or V) are the energy level and effective density of state for each band, and the subscripts C and V represent the conduction and valence band, respectively. We here define the chemical potential in a steady state based on a concept similar to quasi-Fermi level, although the chemical potentials should be discussed under an equilibrium. This is because it may be considered that the intra-band relaxation is fast enough to reach a local equilibrium within each band. Here, we assume that the variation of carrier densities due to illumination and the recombination rate are negligible and we concentrate the discussion on the impact of bias voltage. The applied bias voltage induces a change in the chemical potential, particularly for the minority carriers, compared to the dark condition. Within the *p*-type layer, the behavior of the chemical potential of electrons under a forward applied bias voltage $V$ can be expressed as

$$\mu_e = \mu'_e + qV \qquad (6)$$

where $q$ is the elementary charge and $\mu'_e$ is the chemical potential of electrons in the dark condition. External voltage, which we can manipulate, controls the chemical potential, while what we are primarily interested in is the carrier concentration to evaluate recombination. As $V$ increases, the cell passes through the maximum power point and reaches the open circuit condition. According to Eq. (1), the open circuit voltage ($V_{OC}$) can be interpreted as the voltage that leads to a carrier distribution, which in turn determines a recombination rate that equals the generation rate, resulting in zero current density:

$$j = g - r = 0. \qquad (7)$$

Using Eq. (6), the electron concentration within the *p*-type region under the applied voltage is then given by

$$n_e = n'_e \exp\left(\frac{qV}{kT}\right), \qquad (8)$$

where $n'_e$ is the electron concentration in the dark. Based on Eq. (8), to increase the voltage required to shift the carrier distribution from the dark to open circuit condition, which is $V_{OC}$, reducing $n'_e$ is a possible approach. According to the intrinsic carrier concentration $n_i$, given by

$$n_i^2 = n'_e n'_h = N_C N_V \exp\left(-\frac{E_g}{kT}\right), \qquad (9)$$

where $n'_h$ is the hole concentration in the dark and $E_g$ is the bandgap, there are two methods to for reducing $n'_e$: increasing $n'_h$ and decreasing $n_i$. The former approach can be achieved by increasing the dopant concentration, which corresponds to moving the chemical potential closer to the valence band. For the latter, we consider reducing the effective density of states according to Eq. (9), because the bandgap, which is the other controllable parameter, is strongly related to the absorption edge. By reducing the effective density of states, a given chemical potential can be achieved with a lower minority carrier concentration. The physical meaning is not intuitive, but it can be easily understood through an analogy with an ideal gas. The chemical potential of an ideal gas species *j* is given by

$$\mu_j = \mu_j^\circ + kT \ln \frac{p_j}{p_0}, \qquad (10)$$

where $\mu_j$ is the chemical potential of *j*, $\mu_j^\circ$ is the standard chemical potential for the pure *j* phase, $p_j$ is the partial pressure of *j*, and the $p_0$ is the total pressure. Comparing Eq. (4) with Eq. (10), decreasing the effective density of states corresponds to a decrease in $p_0$, which represents the pressure of the standard state of *j*. Similarly, the effective density of states can be regarded as a parameter that characterizes the standard state of carriers in a semiconductor. A lower density of states implies a higher chemical potential for carriers in standard states. This gain enables the achievement of a higher chemical potential with a lower carrier concentration. The considered principle is observed in the deposition rate difference between MBE and MOCVD. The actual number of atoms reaching the substrate, which corresponds to pi, depends on $p_0$, even though the chemical potential required to produce the same substance is not much different.

Although it has been suggested that a high dopant concentration and a low effective density of states have a potential to enhance $V_{OC}$, how those parameters work on the steady-state carrier distributions throughout solar cells under operating conditions is still unclear. Thus, in the following section, we discuss the steady-state distribution of carrier concentrations and recombination rates under light irradiation and applied bias voltage through a series of device simulations as functions of dopant concentration and effective density of states.

## III. SIMULATION DETAILS

Numerical simulation techniques are useful for investigating steady-state carrier distributions. In this work, simulations were performed using SCAPS-1D, a specialized device simulator for solar cells developed by Burgelman et al. [33]. The calculations are based on continuity equations for electrons and holes as well as Poisson's equation. These coupled equations are solved using the Gummel iteration scheme with Newton-Raphson substeps. We considered a simple drift-diffusion theory and ignored any quantum transport effect.

We primarily conduct simulations by varying the dopant concentration, $N_d$, and the effective densities of states, $N_{eff}$, for hypothetical homo-$pn$ solar cells. $N_d$ denotes the acceptor concentration for $p$-type layers and the donor concentration for $n$-type layers. The effective density of states in the conduction and valence bands were set to the same value, with $N_{eff}$ set to maintain the homo-$pn$ junction. The input parameters are shown in Table I, and a band diagram is depicted in Fig. 1. These parameters were primarily sourced from the literature [36], with particular references to CdTe solar cells. The absorption coefficient, $\alpha$, in the $p$-type region was calculated using the software's "$E_g$-sqrt" model, where $\alpha$ is proportional to the square root of the incident light energy above the bandgap. For the $n$-type region, $\alpha$ was kept constant at 1 cm$^{-1}$ in all wavelengths to ignore the light absorption. In our model, the bandgap of a $p$- and $n$-type layer, $E_g$, was set to 1.4 eV. We also considered a uniform Schockley-Read-Hall (SRH) defect as a primary recombination process [26,27]. The capture cross sections for electrons and holes, the trap density, $N_t$, and trap level were $10^{-13}$ cm$^{-2}$, $2\times10^{14}$ cm$^{-3}$, and 4.98 eV from the vacuum level, respectively. The trap level corresponds to the center of the bandgap in the absorber. The effective carrier lifetime, $\tau_{SRH}$, was 5 ns. Note that the recombination rate (or carrier lifetime, diffusion length, etc.) depends on the carrier concentration at each position in a cell and is a nominal value. The surface recombination velocities for majority and minority carriers were set to $10^7$ and $10^2$ cm s$^{-1}$ respectively, to disregard the diffusion current density, $j_d$, compared to the recombination current density, $r$.

TABLE I. Input parameters for SCAPS-1D.

| Parameters | Values |
| --- | --- |
| Thickness of each layer / μm | 1 |
| Electron affinity, $\chi$ / eV | 4.28 |
| Band gap, $E_g$ / eV | 1.4 |
| Dopant density, $N_d$ / cm$^{-3}$ | $10^{14}$–$10^{20}$ |
| Effective density of states, $N_{eff}$ / cm$^{-3}$ | $10^{14}$–$10^{20}$ |
| Electron mobility / cm$^2$ V$^{-1}$ s$^{-1}$ | $10^2$ |
| Hole mobility / cm$^2$ V$^{-1}$ s$^{-1}$ | $10^2$ |
| Relative dielectric constant | 10.4 |
| Surface recombination velocity for majority carriers / cm s$^{-1}$ | $10^7$ |
| Surface recombination velocity for minority carriers / cm s$^{-1}$ | $10^2$ |
| Trap density, $N_T$ / cm$^{-3}$ | $2\times10^{14}$ |
| Capture cross-section, $\sigma$ / cm$^{-2}$ | $10^{-13}$ |
| Trap level from the vacuum level / eV | 4.98 |

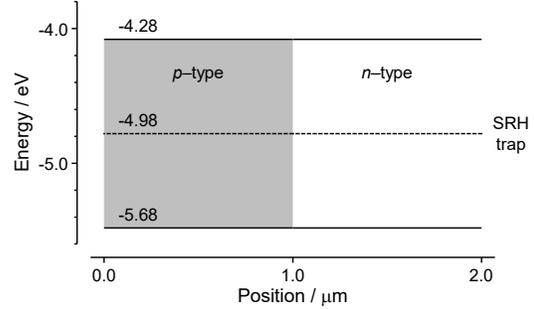

FIG. 1. Schematic band diagram of the homo-$pn$ junction solar cell setup in the device simulator.

## IV. RESULTS AND DISCUSSION
### A. Overview of the simulated photovoltaic performance

First, we provide an overview of how the dopant concentration and effective density of states affect photovoltaic performance before discussing the carrier distribution under specific conditions in the following sub-sections. Figure 2 illustrates the simulated power conversion efficiency (PCE), fill factor (FF), short circuit current density ($J_{SC}$), and open circuit voltage ($V_{OC}$) as functions of dopant concentration in various effective densities of states. As expected in Sec. II, we observe that $V_{OC}$ increases as $N_{eff}$ decreases and $N_d$ increases. Similarly, $FF$ increases with higher $N_d$ and lower $N_{eff}$. On the other hand, $J_{SC}$ reaches a peak value at a specific $N_d$ and increases as $N_{eff}$ decreases. Despite the variation in $J_{SC}$, PCE appears to be strongly influenced by $V_{OC}$. While the simulations provide us with the correlations between these photovoltaic properties and variables, further efforts are necessary to understand the underlying physical origin driving these correlations. To clarify this, as discussed in Sec. II, it is necessary to consider the distribution of recombination rates and carrier concentrations based on the input parameters, such as $N_d$ and $N_{eff}$, and applied conditions. Otherwise, there is a risk of misunderstanding the qualitative insights and physical meaning. Since the most crucial factor for solar cells is PCE, attention is typically focused on the carrier concentration at the maximum power point. However, this concentration is influenced by both illumination and bias voltage, making it more complex. Moreover, the behavior of PCE can easily change depending on input parameters. For example, when $\tau_{SRH}$ is set to 1 ns, PCE exhibits a peak similar to that of $J_{SC}$ in Fig. S1 of the Supplemental Materials (SM) [37]. Rather than focusing on the optimization of PCE, it is more fundamental and insightful for solar cell design to consider the carrier concentration distributions under short circuit and open circuit conditions. By analyzing these carrier concentrations individually, we can gain deeper insights into how carrier distributions are affected by illumination and bias voltage.

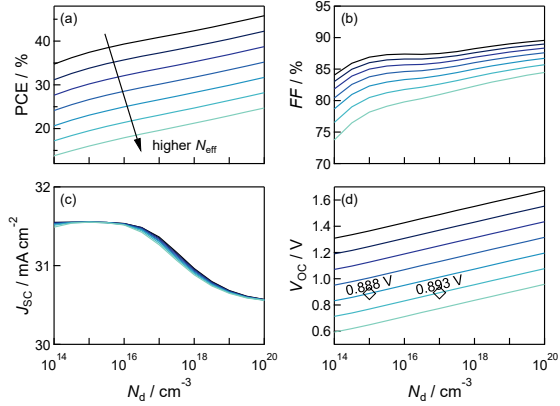

FIG. 2. Simulated (a) power conversion efficiency (PCE), (b) fill factor (FF), (c) short circuit current density ($J_{SC}$), and open circuit voltage ($V_{OC}$) as functions of dopant concentration ($N_d$) for seven different effective densities of states ($N_{eff} = 10^{14}, 10^{15}, 10^{16}, 10^{17}, 10^{18}, 10^{19}, 10^{20}$ cm$^{-3}$). The input parameters are listed in Table I and the SRH lifetime ($\tau_{SRH}$) is set to 5 ns.

### B. Carrier distribution under an open circuit

Figure 3(a) shows the simulated distributions of electron and hole concentrations for three different values of the effective density of states ($N_{eff} = 10^{18}, 10^{19}, 10^{20}$ cm$^{-3}$). The solid lines represent the values under open circuit conditions, while the dashed lines correspond to values under dark conditions. It is evident that a lower $N_{eff}$ results in lower minority carrier concentrations within the bulk region in the dark. Conversely, the carrier distributions under open circuit conditions are independent of $N_{eff}$, producing the same distributions of recombination rates as shown in Fig. 3(b). A lower minority carrier concentration in the dark requires a higher external voltage to achieve carrier distributions satisfying Eq. (7), leading to a higher $V_{OC}$.

Figure 3(c) shows the chemical potential distributions under dark and open circuit conditions. In the dark, although the majority carrier concentrations in each bulk region are the same, their chemical potentials are closer to the bands at lower $N_{eff}$. This is due to the low effective density of states in the valence and conduction bands in $p$- and $n$-type regions, respectively. Similarly, the carrier distributions under open circuit conditions are the same, but the chemical potentials of minority carriers in the bulk region are closer to their bands. As a result, $V_{OC}$, which corresponds to the difference in chemical potentials of minority carriers under the dark and open circuit conditions, increases as $N_{eff}$ decreases.

As shown in Fig. 2(d), $V_{OC}$ changes by 0.119 V for each change in $N_{eff}$ for a given $N_d$. This is consistent with the voltage required to increase the minority carrier concentration by a factor of $10^2$: $kT\ln 10^2 \sim 0.119$ V as derived from Eq. (8), since the effective density of states in both the conduction band and the valence band is changed by an order of magnitude. The influence of the effective density of states on $V_{OC}$ quantitatively matches the expectation described in Sec. II and is also qualitatively confirmed in the literature [38].

Even without improvements in defect properties, a lower $N_{eff}$ can lead to a higher $V_{OC}$. This is based on the principle that increasing the chemical potential of carriers in a standard state allows the same carrier concentrations to result in a higher chemical potential or one closer to the bands. In this way, the relationship between chemical potential and carrier concentration can be altered by controlling the standard state, which in practice corresponds to material selection or band engineering.

Figure 3(d) shows the carrier distributions for three different donor concentrations ($N_d = 10^{15}, 10^{16}, 10^{17}$ cm$^{-3}$) under dark and open circuit conditions. As discussed in Sec. II, a high $N_d$ reduces the minority carrier concentrations within the bulk layer in the dark, leading to a high $V_{\text{OC}}$ as shown in Fig. 2(d). In Fig. 2(d), diamond symbols represent the value with the same minority carrier concentration in the dark. Based on Eqs. (6)–(9), the same minority carrier concentration in the dark would result in the same $V_{\text{OC}}$. However, it is observed that $V_{\text{OC}}$ slightly increases as $N_\text{d}$ increases under this condition. This is because the distribution of recombination rates under open circuit conditions depends on $N_\text{d}$ as shown in Fig. 3(e).

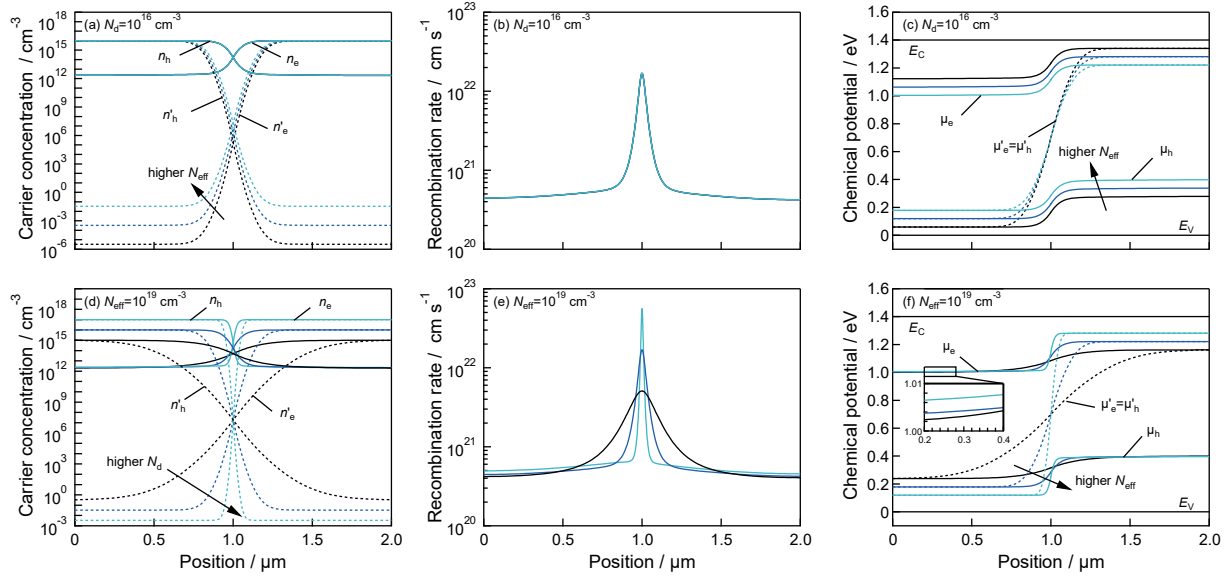

FIG. 3. Distributions of (a) carrier concentrations, (b) recombination rates, and (c) chemical potentials simulated for different effective densities of states ($N_{\text{eff}} = 10^{18}$, $10^{19}$, $10^{20}$ cm$^{-3}$) with dopant concentration, $N_\text{d}$, of $10^{16}$ cm$^{-3}$. Distributions of (d) carrier concentrations, (e) recombination rates, and (f) chemical potentials simulated for different dopant concentrations ($N_\text{d} = 10^{15}$, $10^{16}$, $10^{17}$ cm$^{-3}$) $N_{\text{eff}}$ of $10^{16}$ cm$^{-3}$. Solid and dashed lines denote distributions calculated under open circuit and dark conditions, respectively.

Under open circuit conditions, the integral of recombination rates must be the same as that of generation rates, according to Eq. (7). The proportion of the depletion layer under open circuit conditions depends on $N_\text{d}$, and the recombination rates in this region also change. On the other hand, the recombination rates within the bulk also change with $N_\text{d}$, while the total recombination rate remains equal to the total generation rate. In the bulk layer, a high recombination rate indicates a high minority carrier concentration, making its chemical potential high. This corresponds to an increase on the left-hand side of Eq. (6), resulting in a rise in $V_{\text{OC}}$. In this system, in which SRH defects are uniformly distributed, as $N_\text{d}$ increases, the recombination rate peak in the depletion layer rises and its width becomes narrow. Consequently, the higher $N_\text{d}$ slightly increases the minority carrier concentration in the bulk layer, leading to an increase in its chemical potential as confirmed in the inset of Fig. 3(f). This is the reason why $V_{\text{OC}}$ slightly increases with an increase in $N_\text{d}$ increase under the condition in which the minority carrier concentration in the dark is the same. By adjusting $N_\text{d}$, we can control the distribution of recombination rates under open circuit conditions and increase the chemical potential of minority carriers within the bulk layers, thereby improving $V_{\text{OC}}$. This concept corresponds to distributing the recombination rate, equal to the total generation rate, as much as possible within the bulk region.

It should be noted that the optimum $N_\text{d}$ depends on the distribution of defects. As shown in Fig. S2 (d) in the SM [37], the introduction of an interface defect can cause a lower $N_\text{d}$ to result in a higher $V_{\text{OC}}$. This contrasts with the previous findings. In actual solar cells, an optimal $N_\text{d}$ exists as a balance between increased interface recombination with rising $N_\text{d}$ and decreased minority carrier concentration in the dark.

### C. Carrier distributions under a short circuit

Initially, we show that the transition from the carrier distribution in the dark to that under short circuit conditions is dominated by drift, despite both drift and diffusion being considered in the simulation [33]. We worked on conceptual conditions in which the driving force for drift is the same, while that for diffusion is not. As tabulated in the Table II, we adjusted $N_{\text{eff}}$ and dielectric constant, ε, for five different values of $N_\text{d}$ (=$10^{14}$, $10^{15}$, $10^{16}$, $10^{17}$, $10^{18}$

TABLE II. Input parameters for Fig. 4

| $N_d$ / cm$^{-3}$ | $N_{eff}$ / cm$^{-3}$ | ε |
|---|---|---|
| $10^{14}$ | $10^{16}$ | 0.104 |
| $10^{15}$ | $10^{17}$ | 1.04 |
| $10^{16}$ | $10^{18}$ | 10.4 |
| $10^{17}$ | $10^{19}$ | 104 |
| $10^{18}$ | $10^{20}$ | 1040 |

cm$^{-3}$) to ensure that the chemical potential distributions in the dark are the same. The carrier concentration and chemical potential in the dark are shown in Figs 4(a) and 4(b), respectively. All other conditions remain consistent with those in Table I. Since the electric and chemical potentials are balanced in the dark, the electric potentials are independent of $N_d$, leading to identical electric fields in the dark as shown in Fig. 4(c). When these cells are illuminated, the driving forces of photogenerated carrier movement via drift will be the same. Conversely, differences in carrier distribution in the dark would modify the driving force for diffusion. For example, a high $N_d$ would enhance the diffusion of photogenerated holes in the p-type layer to the n-type layer due to the steep negative gradient of the hole concentration. Strictly speaking, the driving force for diffusion is the gradient of the chemical potential; however, the same argument applies here because the generation rates are the same, but $N_{eff}$ differs, resulting in different chemical potential gradients immediately after illumination.

Figure 4(d) shows the carrier concentrations under short circuit conditions. The distributions of minority carriers, which are electrons in the range of 0–1.0 μm and holes in the range 1.0–2.0 μm, are almost the same. In the vicinity of the pn junction, the photogenerated hole and electron concentrations in the p- and n-type regions, respectively, also converge to the same value as they approach the junction. This suggests that the photogenerated carrier distribution under short circuit conditions is determined by the drift, which is driven by the electric field, and the contribution of diffusion is negligible. A closer look at Fig. 4(d) reveals that the carrier concentration in each layer (in the ranges of 0 to 0.5 and 1.5 to 2.0 μm) is slightly lower in cases with lower $N_d$. This is due to the slight enhancement of the electric field in the region by photogenerated carriers, as shown in Fig. 3S in the SM [37]. Consequently, by considering conceptual conditions, it has been elucidated that the change in carrier concentration due to illumination in the solar cell is dominated by drift, with the electric field acting as the driving force. Since the steady-state distribution of carrier concentrations determines that of recombination rates, and thus $r$, $j$ is characterized by the electric field. In this context, it is concluded that the driving force of carrier separation in a solar cell is the internal electric field. Several studies have suggested that carrier separation is governed by the internal electric field or the gradient of the electrochemical potential [34,36,39,40]. However, these perspectives differ from the approach taken in this study. The literature assumes transient carrier transport such as the behavior of additional carriers introduced into a steady-state cell. This assumption contradicts the fact that an operating solar cell is inherently in a steady state. In this condition, there is no temporal change in the carrier distribution, and the net current density, including carrier flow, generation rate, and recombination rate, remains constant throughout the solar cell. As a result, there is no net driving force for carrier movement. The driving force for carrier separation refers to the force that acts to bring the solar cell from the moment immediately after illumination into a steady state under illumination, and this force is the electric field.

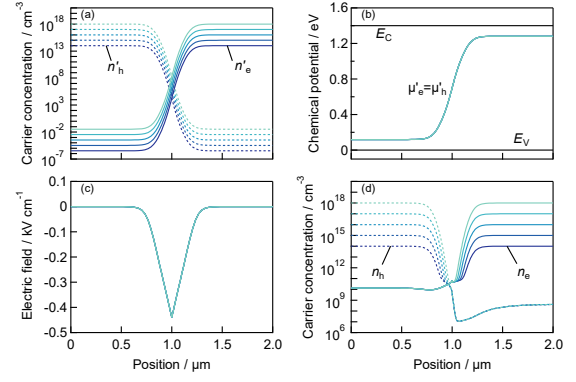

FIG. 4. (a) Carrier concentrations, (b) chemical potentials, and (c) electric fields in the dark simulated for different values of $N_d$ (=$10^{14}$, $10^{15}$, $10^{16}$, $10^{17}$, $10^{18}$ cm$^{-3}$) with adjustments of $N_{eff}$ and ε. The varied input parameters are listed in Table II, and the others are the same as those in Table I. (d) Carrier distributions under short circuit conditions.

Incidentally, the mobility and absorption coefficient also affect the transient behaviors; however, they were treated as fixed values in this study. Higher mobility promotes drift, and higher generation rates lead to higher minority carrier concentrations. The details are shown in Fig. S4 in the SM [37].

In light of the above discussion, we investigated the impact of $N_d$ and $N_{eff}$ on the carrier distribution under short circuit conditions. Fig. 5(a) shows the carrier distributions under short circuit conditions. Calculations were performed for three different values of effective density of states ($N_{eff} = 10^{18}$, $10^{19}$, $10^{20}$ cm$^{-3}$) with $N_d$ fixed at $10^{16}$ cm$^{-3}$. The minority carrier concentrations are slightly lower in lower $N_{eff}$. This behavior is consistent with the electric fields, as shown on a logarithmic scale in Fig. 5(b). As explained in Sec. II and IV-B, a low $N_{\text{eff}}$ raises the chemical potential of electrons (or decreases that of holes), which increases the built-in potential and strengthens the electric fields. Therefore, a lower $N_{\text{eff}}$ results in lower minority carrier concentrations due to the higher electric field. Figure 5(c) shows the recombination rate distributions. As $N_{\text{eff}}$ decreases, the recombination rate at each position decreases due to the low minority carrier concentration. This explains the fact that higher $N_{\text{eff}}$ results in higher $J_{\text{SC}}$, as shown in Fig. 2(c).

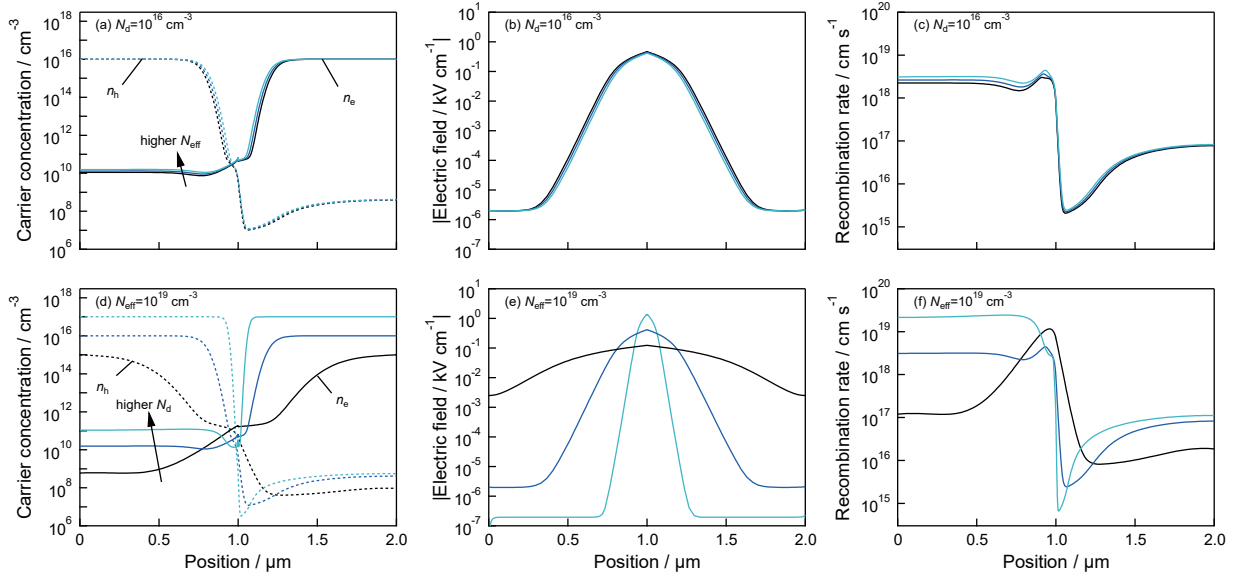

FIG. 5. (a) Carrier concentrations, (b) electric fields, and (c) recombination rates under short circuit conditions simulated for different values of $N_{\text{eff}}$ (=$10^{17}$, $10^{18}$, $10^{19}$ cm$^{-3}$) with $N_d$ fixed at $10^{16}$ cm$^{-3}$. (d) Carrier distributions, (e) electric fields, and (f) recombination rates under short circuit conditions simulated for different values of $N_d$ (=$10^{15}$, $10^{16}$, $10^{17}$ cm$^{-3}$) with constant $N_{\text{eff}}$ of $10^{19}$ cm$^{-3}$.

Figure 5(d) shows the carrier concentrations under short circuit conditions with $N_d$ varying from $10^{15}$ to $10^{17}$ cm$^{-3}$ while keeping $N_{\text{eff}}$ constant at $10^{19}$ cm$^{-3}$. The relative magnitudes of minority carrier concentrations for different values of $N_d$ depend on the position. In the bulk regions, a low $N_d$ leads to a low minority carrier concentration. Conversely, for example, in the vicinity of the pn junction on the p-type side, the electron concentration decreases as $N_d$ increases and is inverted when $N_d$ is $10^{15}$ and $10^{16}$ cm$^{-3}$. These scaling relationships are consistent with our previous findings. Figure 5(e) shows the electric field distributions for different $N_d$ values. A lower $N_d$ strengthens the electric field in the bulk region but weakens it near the pn junction. These variations in electric fields affect the distribution of recombination rates shown in Fig. 5(f) via carrier distributions. For example, a lower $N_d$ can reduce the recombination rate in the bulk region but promote it around the junction. This trade-off results in an optimal $N_d$ value that maximizes $J_{\text{SC}}$, as shown in Fig. 2(c). The optimal $N_d$ for $J_{\text{SC}}$ also depends on the defect distribution, as in the case for $V_{\text{OC}}$. As shown in Fig. S1(c) in the SM [37], the optimal $N_d$ shifts to a lower value when the SRH lifetime is 1 ns. This is due to the increased significance of recombination within the layers. Furthermore, when only interface defects are introduced, a higher $N_d$ leads to a higher $J_{\text{SC}}$, as shown in Fig. S2(c) in the SM [37]. This occurs because the enhanced electric field at the interface reduces recombination.

In conclusion, the steady-state carrier distribution under short circuit conditions can be controlled by the electric field distribution. Practical methods for controlling the electric field may include not only $N_d$ and $N_{\text{eff}}$ but also the dielectric constant, which was used for conceptual cells at the beginning of this section. Optimizing the electric field distribution depending on defect characteristics is important for improving $J_{\text{SC}}$, which is similar to control of the distribution of recombination rates under open circuit conditions, as discussed in Sec. IV-B. Under short circuit conditions, the suppression of recombination due to the reduction of carrier concentration is literally field-effect passivation.

## V. CONCLUSIONS

In this study, we investigated the effects of illumination and bias voltage on steady-state distributions of carrier concentrations and recombination using SCAPS-1D. Our aim was to determine whether the reduction of carrier concentrations can be an effective passivation throughout the entire solar cells. The current density is defined based on Eq. (1) as the difference between the integrals of generation and recombination rates, with the recombination rates being determined by carrier concentrations. Furthermore, we discussed carrier distributions by carefully separating the chemical potential from the electric potential.

We based our study on the principle that the current is determined by the carrier concentration, while bias voltage changes the chemical potential. By controlling the standard states of carriers, lower carrier concentrations are needed to achieve a given chemical potential, leading to an increase in $V_{OC}$. Controlling the standard state is possible not only through material selection, such as the effect of $N_{eff}$ in this study, but also through band engineering or the utilization of hetero-junctions. Additionally, we found that optimizing the distribution of recombination rates depending on the defect structure under open circuit conditions is crucial. Based on the principle that the total recombination rate equals the total generation rate under this condition, increasing the recombination rate in the bulk region as much as possible provides a way for improving $V_{OC}$.

By considering conceptual cells, we showed that the carrier distribution under short circuit conditions is governed by the electric field. Since the carrier concentration determines the current, it is concluded that the electric field is the driving force for carrier separation in solar cells. Optimizing the internal electric field distribution depending on defect characteristics and distribution is expected to improve $J_{SC}$.

It is important to note that the capture cross-sections of SRH defects for electrons and holes differ [41], and trap levels are often not located at the center of the bandgap. These characteristics influence the optimal carrier distributions under both open and short circuit conditions. Additionally, when chemical passivation successfully reduces defect density in the bulk or at the interface, further efficiency improvements can be expected by adjusting the carrier concentration and electric field to the defect structure of that situation

This paper has shown that focusing on the precise distribution of carrier concentrations, chemical potentials, and electric fields under various conditions provides comprehensive insights into reducing recombination rates in solar cells. To achieve higher conversion efficiency, it is essential to consider the carrier distribution at the maximum power point, based on the dependencies of carrier concentration and recombination distribution on illumination and bias voltage, as demonstrated in this study.

## ACKNOWLEDGMENTS

This work was supported by JST SPRING (No. JPMJSP2110), JSPS KAKENHI (Nos. 23K26432 and 24KJ1392), and Collaborative Research Project of Materials and Structures Laboratory, Institute of Integrated Research, Institute of Science Tokyo.

[1] M. Wolf, Proc. IEEE **51**, 674 (1963).
[2] J. Mandelkorn and J. H. Lamneck, Solar Cells **29**, 121 (1990).
[3] G. Du, B. Chen, N. Chen, and R. Hu, IEEE Electron Device Lett. **33**, 573 (2012).
[4] R. R. King, E. W. Thomas, W. B. Carter, and A. Rohatgi, in *Proceedings of 21st IEEE Photovoltaic Specialists Conference* (New York, 1991), p. 229, vol.1.
[5] Z. T. Kuznicki, Sol. Energy Mater. Sol. Cells **31**, 383 (1993).
[6] A. W. Blakers, A. Wang, A. M. Milne, J. Zhao, and M. A. Green, Appl. Phys. Lett. **55**, 1363 (1989).
[7] M. A. Green, A. W. Blakers, J. Shi, E. M. Keller, and S. R. Wenham, IEEE Trans. Electron Devices **31**, 679 (1984).
[8] M. A. Green, A. W. Blakers, J. Zhao, A. M. Milne, A. Wang, and X. Dai, IEEE Trans. Electron Devices **37**, 331 (1990).
[9] A. G. Aberle, Prog. Photovolt.: Res. Appl. **8**, 473 (2000).
[10] R. J. Schwartz and M. D. Lammert, in *Proceedings of Silicon Solar Cells for High Concentration Applications*, in *1975 International Electron Devices Meeting* (IEEE, New York, 1975), p. 350.
[11] R. A. Sinton, Young Kwark, J. Y. Gan, and R. M. Swanson, IEEE Electron Device Lett. **7**, 567 (1986).
[12] R. R. King, R. A. Sinton, and R. M. Swanson, Appl. Phys. Lett. **54**, 1460 (1989).
[13] K. Yoshikawa, H. Kawasaki, W. Yoshida, T. Irie, K. Konishi, K. Nakano, T. Uto, D. Adachi, M. Kanematsu, and H. Uzu *et al*., Nat. Energy **2**, 17032 (2017).
[14] E. V. Kerschaver and G. Beaucarne, Prog. Photovolt.: Res. Appl. **14**, 107 (2006).
[15] M. Taguchi, A. Yano, S. Tohoda, K. Matsuyama, Y. Nakamura, T. Nishiwaki, K. Fujita, and E. Maruyama, IEEE J. Photovolt. **4**, 96 (2014).
[16] D. Adachi, J. L. Hernández, and K. Yamamoto, Appl. Phys. Lett. **107**, 233506 (2015).
[17] M. Taguchi, K. Kawamoto, S. Tsuge, T. Baba, H. Sakata, M. Morizane, K. Uchihashi, N. Nakamura, S. Kiyama, and O. Oota, Prog. Photovolt.: Res. Appl. **8**, 503 (2000).
[18] F. Feldmann, M. B. C. Reichel, M. Hermle, and S. W. Glunz, in *Proceedings of 28th European Photovoltaic Solar Energy Conference and Exhibition*, A. Mine and A. J. Waldau (WIP-Renewable Energies, Munich, 2013), p. 988.
[19] S. W. Glunz, F. Feldmann, A. Richter, M. Bivour, C. Reichel, H. Steinkemper, J. Benick, and M. Hermle, in *Proceedings of 31st European Photovoltaic Solar Energy Conference and Exhibition (EU PVSEC 2015)*,

S. Rinck, N. Talyor, and P. Helm (WIP-Renewable Energies, Munich, 2015) p. 259.

[20] D. K. Ghosh, S. Bose, G. Das, S. Acharyya, A. Nandi, S. Mukhopadhyay, and A. Sengupta, Surf. Interfaces **30**, 101917 (2022).

[21] E. Yablonovitch, D. L. Allara, C. C. Chang, T. Gmitter, and T. B. Bright, Phys. Rev. Lett. **57**, 249 (1986).

[22] T. Lauinger, J. Schmidt, A. G. Aberle, and R. Hezel, Appl. Phys. Lett. **68**, 1232 (1996).

[23] M. J. Kerr, J. Schmidt, A. Cuevas, and J. H. Bultman, J. Appl. Physs. **89**, 3821 (2001).

[24] J. I. Pankove and M. L. Tarng, Appl. Phys. Lett. **34**, 156 (1979).

[25] S. Olibet, E. Vallat-Sauvain, and C. Ballif, Phys. Rev. B **76**, 035326 (2007).

[26] W. Shockley and W. T. Read, Phys. Rev. **87**, 835 (1952).

[27] R. N. Hall, Phys. Rev. **87**, 387 (1952).

[28] O. Lundberg, M. Edoff, and L. Stolt, Thin Solid Films **480–481**, 520 (2005).

[29] B. Barman and P. K. Kalita, Sol. Energy **216**, 329 (2021).

[30] P. D. Demoulin, M. S. Lundstrom, and R. J. Schwartz, Solar Cells **20**, 229 (1987).

[31] N. Amin, M. A. Matin, M. M. Aliyu, M. A. Alghoul, M. R. Karim, and K. Sopian, Int. J. of Photoenergy **2010**, 1 (2010).

[32] M. Mamta, R. Kumari, K. K. Maurya, and V. N. Singh, Energy Technol. **11**, 2201522 (2023).

[33] M. Burgelman, P. Nollet, and S. Degrave, Thin Solid Films **361–362**, 527 (2000).

[34] U. Wurfel, A. Cuevas, and P. Wurfel, IEEE J. Photovolt. **5**, 461 (2015).

[35] A. Cuevas, Y. Wan, D. Yan, C. Samundsett, T. Allen, X. Zhang, J. Cui, and J. Bullock, Sol. Energy Mat. Sol. Cells **184**, 38 (2018).

[36] K. O. Hara and N. Usami, J. Appl. Phys. **114**, 153101 (2013).

[37] See Supplemental Material at [url] for influences of defect characteristics on photovoltaic parameters, detailed results of simulations under conceptual conditions where the driving force for drift is the same, while that for diffusion is not, and effects of mobilities and absorption coefficients on distributions of carrier concentration and recombination rate.

[38] T. Kirchartz and U. Rau, Sustain. Energy Fuels **2**, 1550 (2018).

[39] A. Cuevas and D. Yan, IEEE J. Photovolt. **3**, 916 (2013).

[40] B. Lipovšek, F. Smole, M. Topič, I Humar, A. R. Sinigoj, AIP Adv. **9**, 055026 (2019).

[41] E. Yablonovitch, R. M. Swanson, W. D. Eades, B. R. Weinberger, Appl. Phys. Lett. **48**, 245 (1986)

# Supplemental material for: Steady-state distributions of carrier concentration and recombination rates in a solar cell under operating conditions

Isshin Sumiyoshi,[1,*] Yoshitaro Nose,[1,†]

[1]*Department of Materials Science and Engineering, Kyoto University, Sakyo-ku, Kyoto 606-8501, Japan*
(Dated: October 31st, 2024)

SUMPELENTAL MATERIAL

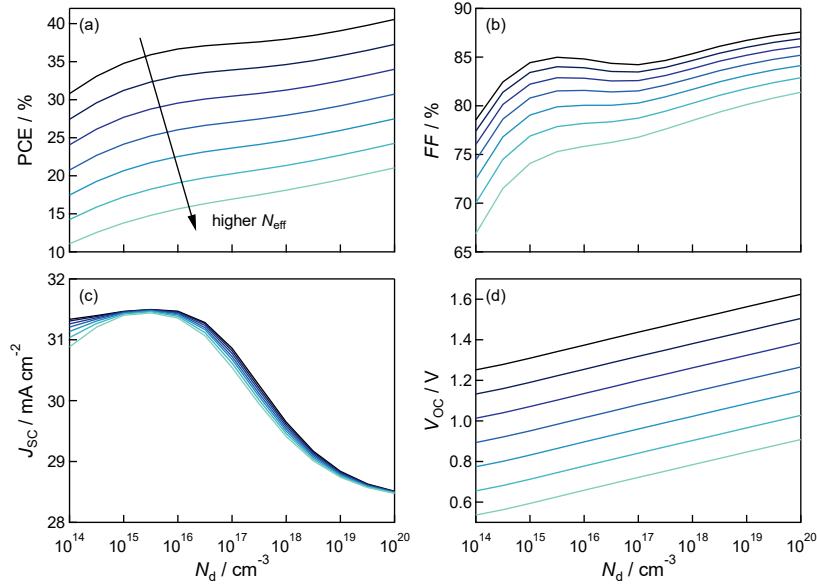

FIG. S1. Simulated (a) power conversion efficiency (PCE), (b) fill factor ($FF$), (c) short circuit current density ($J_{SC}$), and open circuit voltage ($V_{OC}$) as functions of dopant concentration ($N_d$) for seven different effective densities of states ($N_{eff} = 10^{14}, 10^{15}, 10^{16}, 10^{17}, 10^{18}, 10^{19}, 10^{20}$ cm$^{-3}$). The input parameters, except for trap density ($N_T$), are the same as those listed in Table I, and the SRH lifetime ($\tau_{SRH}$) is 1 ns.

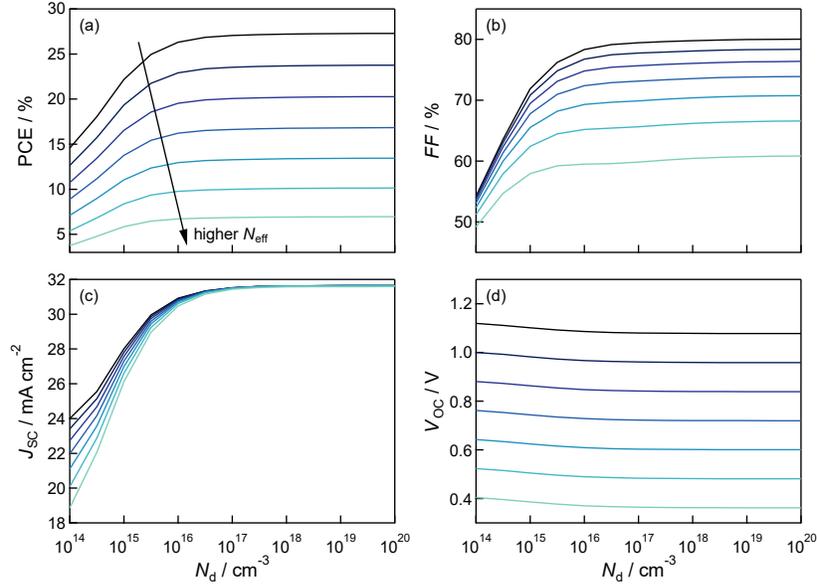

FIG. S2. Simulated (a) power conversion efficiency (PCE), (b) fill factor (FF), (c) short circuit current density ($J_{SC}$), and (d) open circuit voltage ($V_{OC}$) as functions of dopant concentration ($N_d$) for seven different effective densities of states ($N_{eff}$ = $10^{14}$, $10^{15}$, $10^{16}$, $10^{17}$, $10^{18}$, $10^{19}$, $10^{20}$ cm$^{-3}$) when an interface defect is considered. The trap level, capture cross sections, and trap density were set to the middle of the semiconductor bandgap, $10^{-13}$ cm$^2$, and $10^{12}$ cm$^{-2}$, respectively.

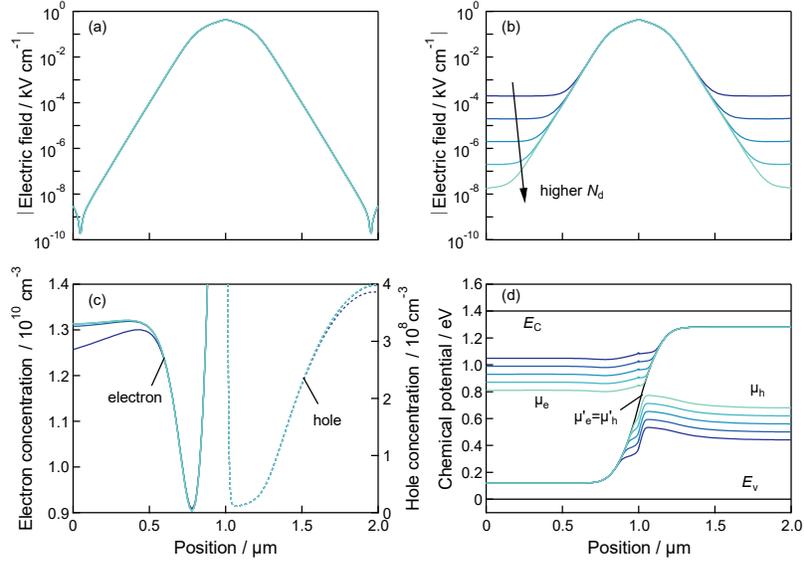

FIG. S3. Distributions of electric field in the dark (a) and under short circuit conditions (b) for different values of $N_d$ (=$10^{14}$, $10^{15}$, $10^{16}$, $10^{17}$, $10^{18}$ cm$^{-3}$) with adjustments of $N_{eff}$ and $\varepsilon$. Values are shown on a logarithmic scale. (c) Electron and hole concentrations under short circuit conditions corresponding to the simulated results in FIG. S3(b). In a low $N_d$, the excess minority carrier concentration due to illumination is relatively high compared to $N_d$ itself. The influence of charge of excess carriers on Poisson's equation cannot be neglected. (d) Chemical potential distributions in the dark and under short circuit conditions.

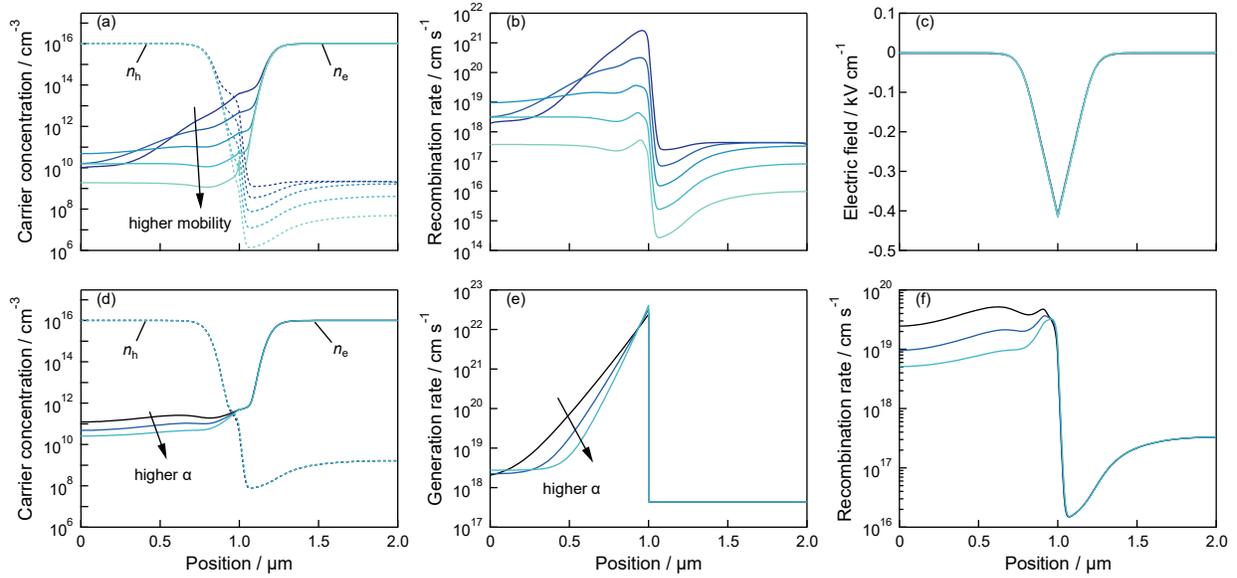

FIG. S4. (a) Carrier concentrations, (b) recombination rates, and (c) electric fields under short circuit conditions simulated for different mobilities (=$10^{-1}$, $10^0$, $10^1$, $10^2$, $10^3$ V cm$^{-1}$ s$^{-1}$). High mobility results in a low minority carrier concentration and thus a low recombination rate. Electron concentrations for mobilities of $10^{-1}$ and $10^0$ V cm$^{-1}$ s$^{-1}$ in the range of 0 to 0.7 μm show non-systematic behavior due to surface recombination at the left contact. According to the drift-diffusion theory, mobility is the coefficient of the driving forces, which explains the variation in minority carrier concentrations despite the constant electric field, as shown in FIG. S4(c). FIG. S4(d), (e), and (f) show the carrier concentrations, generation rates, and recombination rates calculated for scenarios where the absorption coefficient α, in the absorbing layer is multiplied by 0.75, 1, and 1.25. The higher α is, the lower is the electron concentration and the lower is the recombination rate within the *p*-type region. This is because a high α indicates a lower generation rate within the layer, as shown in FIG. S4(e). Generally, an absorbing layer with a low α needs to be thicker to absorb sufficient light, leading to an increase in the total amount of recombination. However, our results show that a low α induces a high recombination rate within the absorbing layer. Note that electric fields are almost identical in this case as well. Moreover, hole concentrations within the *n*-type region are equal. This further supports our argument that the driving force behind the carrier movement from dark to short circuit conditions is the electric field.